\documentclass[fleqn,usenatbib]{mnras}

\usepackage{newtxtext,newtxmath}
\usepackage{hyperref}
\usepackage[T1]{fontenc}

\DeclareRobustCommand{\VAN}[3]{#2}
\let\VANthebibliography\thebibliography
\def\thebibliography{\DeclareRobustCommand{\VAN}[3]{##3}\VANthebibliography}

\usepackage{graphicx}	% Including figure files
\usepackage{amsmath}	% Advanced maths commands
\title[Kinematic substructure in star clusters]{Kinematic substructure in star clusters constrains star cluster formation}

\author[B. Arnold et al.]{
Becky Arnold$^{1}$\thanks{E-mail: r.j.arnold.uk@gmail.com}
and Nicholas J. Wright,$^{1}$
\\
$^{1}$Lennard-Jones Laboratory, Keele University, Newcastle, ST5 5BB, UK\\
}

\date{Accepted XXX. Received YYY; in original form ZZZ}

\pubyear{2023}

\begin{document}
\label{firstpage}
\pagerange{\pageref{firstpage}--\pageref{lastpage}}
\maketitle

% Abstract of the paper
\begin{abstract}
The spatial-kinematic structure of 48 young star clusters and associations is investigated. Moran's $I$ statistic is used to quantify the degree of kinematic substructure in each region, and the results are compared to those expected assuming the hierarchical or monolithic models of star cluster formation. Of the observed regions, 39 are found to have significant kinematic substructure, such that they are compatible with the hierarchical model and incompatible with the monolithic model. This includes multiple regions whose $Q$ parameter shows the region to be centrally concentrated and clustered. The remaining nine are compatible with both models. From this it is concluded that the kinematic substructure of the observed star clusters represents strong evidence in favour the hierarchical model of star cluster formation over the monolithic model.
\end{abstract}

% Select between one and six entries from the list of approved keywords.
% Don't make up new ones.

%%%%%%%%%%%%%%%%%%%%%%%%%%%%%%%%%%%%%%%%%%%%%%%%%%

%%%%%%%%%%%%%%%%% BODY OF PAPER %%%%%%%%%%%%%%%%%%

\section{Introduction} \label{intro}

Most stars, including the Sun, were born in clusters of hundreds to millions of stars \citep{Elmegreen00, Lada03}. The high stellar densities and extreme radiation fields in these clusters has a major impact on the star and planet formation process \citep[e.g.,][]{Bonnell01,Adams06}. As such, the question of how stars form is virtually inseparable from the question of how star clusters form.

There are two models of star cluster formation, the hierarchical mergers model and the monolithic model. In the former, star clusters are thought to form from mergers between substructures that form over a larger volume, but are brought together by gravity (\citealp*{Vazquez-Semadeni17};  \citealp*{Rodriguez20}; \citealp{Sabbi22}). In the latter, it is thought that star clusters form monolithically from approximately spherical, dense concentrations of gas, which must collapse first (\citealp*{Kroupa01}; \citealp{Krumholz07, Farias23}). The models effectively differ in whether material is brought together into a dense configuration before (for the monolithic model) or after (for the hierarchical model) star formation occurs.

These two models make contrasting predictions about the spatial and kinematic structure of young star-forming regions \citep{Banerjee14,Longmore14,Vazquez-Semadeni17}. These differing predictions can be used to try to answer the question of which model is best supported by observational evidence, by comparing them to the structure of real star-forming regions. This goal is obstructed by the fact that spatial structure (which has historically been easier to observe than kinematic structure) is erased quickly as clusters evolve (\citealt{Goodwin04}; \citealt*{Arnold22}). Kinematic structure, however, may yet be of use to delineate between these models observationally. 

This paper follows on from \citet{Arnold22} (hereafter A22), which makes use of $N$-body simulations of star cluster formation. It shows that kinematic substructure can be identified and quantified using Moran's $I$ statistic \citep{Moran50} (a multidisciplinary statistical method used to quantify structure of given properties) even after spatial substructure has been erased. This statistic was found by A22 to be reliable even in the face of highly imperfect data (e.g. large uncertainties, high contamination, low completeness). 

In A22 the evolution of Moran's $I$ in simulations based on the monolithic and hierarchical models of star formation is compared. The initial conditions for the monolithic cases are generated as Plummer spheres \citep{Plummer11}, resulting in centrally-concentrated, smooth, and spherical initial conditions without initial kinematic substructure. A22 finds that Moran's $I$ for such regions is zero within a very small margin or error, and remains so for at least 10 Myr unless the region is expanding. In that case Moran's $I$ increases from $t = 0$ before plateauing.

The initial conditions for simulations based on the hierarchical model are generated using box fractals, in a method analogous to that used in \citet{Allison10}, \citet{DaffernPowell20} and \citet{BlaylockSquibbs22} which results in spatial and kinematic substructure. A22 finds that Moran's $I$ is significantly above zero at $t = 0$ in these cases, and remains so for up to 10 Myr, even though it decreases over time. Expanding regions are an exception. Like in the monolithic case Moran's $I$ increases and plateaus over time, however the hierarchical expanding simulations maintain higher Moran's $I$ values than their monolithic counterparts.

The aim of this paper is to use Moran’s $I$ statistic to measure kinematic substructure in young star clusters. We then aim to determine if the observed levels are consistent with those expected by the hierarchical merger or monolithic star cluster formation models. To do this, data is gathered for 48 currently or recently star-forming regions, and their spatial-kinematic structure is quantified using Moran's $I$ statistic. These values are then compared to the predictions of A22. The relationship between the Moran's $I$ values of different regions and their other properties, such as their spatial structure, is also examined.

In Section \ref{methods} the statistical methods used to analyse these data are described, and in section \ref{sec_obs_data} the observational data is outlined. Section \ref{sec_results} presents the results of the analysis, and section \ref{sec_conclusions} summarises the conclusions drawn from those results.

\section{Methods} \label{methods}

In this section, the statistical methods used to assess the spatial and kinematic structure of the observed star-forming regions are outlined.

\subsection{Moran's $I$ statistic} \label{method_morans}

The principle statistic used in this paper to assess the spatial-kinematic structure of star clusters is Moran's $I$ \citep{Moran50}. This statistical tool is applied to spatial data of a given property in order to investigate the degree of structure in the spatial arrangement of that property. It is used in a wide variety of fields from epidemiology \citep[e.g.,][who investigated the spatial distribution of Covid-19 cases]{Isazade23} to waste management \citep[e.g.,][who investigated how re-use rates vary across the Netherlands]{Tusi22} and geography \citep*[e.g.,][who investigatesd the spatial distribution of small manufacturing firms]{Banasick09}.

Moran's $I$ for $N$ data points is calculated as follows. First, a weight, $w_{ij}$, is calculated for every possible pair of data points, $i$ and $j$, where $i \neq j$. The weight is the inverse of the physical distance between the data points. Once this has been done the Moran's $I$ statistic for a given parameter $\mu$ can be calculated as follows:

\begin{equation} \label{morans_i_eqn}
  I(\mu) = \frac{N}{\sum_{i \ne j}{w_{ij}}}\frac{\sum_{i \ne j}{w_{ij}(\mu_i - \overline{\mu})(\mu_j - \overline{\mu})}}{\sum_{i}{(\mu_i - \overline{\mu})^2}}.
\end{equation}

The resulting Moran's $I$ statistic varies between 1 and -1. Positive values indicate the presence of spatial correlation in the distribution of $\mu$, i.e. similar values of $\mu$ tend to be found in close proximity to one another. Negative values of Moran's $I$ indicate the presence of spatial anti-correlation, i.e. dissimilar values of $\mu$ tend to be found in close proximity to one another. If $\mu$ has no spatial structure, the expected value of Moran's $I$ is near zero\footnote{Strictly speaking, the expected Moran's $I$ assuming no structure is $-N^{-1}$, but given the large number of stars included in the samples used in this work, and the uncertainties on the kinematic data, the approximation of $E$($I$) to zero is valid.}.

In A22 it is demonstrated that Moran's $I$ statistic can be applied to star-forming regions to quantify the spatial correlation of their stellar kinematics. A number of different kinematic properties can take the role of $\mu$, enabling this statistic to study a variety of different aspects of the spatial-kinematic structure in a data set. The parameters used in A22 are velocity in the $x$ direction, $v_{x}$, velocity in the $y$ direction, $v_{y}$, and speed $s$. The calculated Moran's $I$ values of these parameters are referred to as $I$($v_x$), $I$($v_y$), and $I$($s$) respectively. 

To reduce the impact of the frame of reference choice and produce a less noisy metric $I$($v_x$) and $I$($v_y$) are averaged to produce $I$($v_{2D}$). For real rather than simulated data, proper motion velocities ($I$($\mu_{{\alpha}{\ast}}$) and $I$($\mu_{\delta}$)) are used in the place of $I$($v_x$) and $I$($v_y$). Other parameterisations used in this paper are $I$($v_{r}$), where $v_{r}$ is radial velocity, $I$($v_{3D}$), which is analogous to $I$($v_{2D}$) but incorporates $v_{r}$, $I$($s_{2D}$), where $s_{2D}$ is the stellar speed considering only proper motions, and $I$($s_{3D}$) is the same but also incorporating $v_{r}$. Note that because $s$ is always positive $I$($s_{2D}$) and $I$($s_{3D}$) quantify quite different aspects of kinematic structure to $I$($v_{2D}$) and $I$($v_{3D}$). This is expanded upon in section \ref{sec_expand}.

\subsection{The LISA statistic} \label{method_LISA}

LISA \citep[Local Indicators of Spatial Association,][]{Anselin95} is a method of decomposing global statistics such as Moran's $I$. This statistic is calculated for each data point individually (rather than for the region as a whole), and quantifies how much they contribute to the overall global statistic. In this context it is useful for highlighting which stars or subregions are exhibiting the most or least substructure, and aids understanding of the overall dynamics of a region.

As with Moran's $I$, increasingly positive values of the LISA statistic indicate the data point's net contribution is to increase the observed degree of substructure in the region. Likewise, negative values of the LISA statistic indicate the data point contributes negatively. A dissimilarity between Moran's $I$ and the LISA statistic is that the LISA statistic is not confined to a specific range, whereas Moran's $I$ can only exist between -1 and 1. 

The LISA statistic for Moran's $I$ is:

\begin{equation} \label{LISA_eqn}
  I_i = (\mu_i - \overline{\mu})\sum_{i \ne j}{w_{ij}(\mu_j - \overline{\mu})}.
\end{equation}

\noindent for a given data point $i$ and a given parameter $\mu$.  

\subsection{The $Q$ parameter}

The $Q$ parameter \citep{Cartwright04} is a measure of the 2D (projected) spatial distribution of stars that is commonly used to quantify the degree of substructure in star clusters \citep[e.g.,][]{Wright14}. The $Q$ parameter is defined as $Q = \bar{m} / \bar{s}$, where $\bar{m}$ is the normalised mean edge length of the minimum spanning tree of all stars in a cluster, and $\bar{s}$ is the normalised mean separation between the same sample of stars. Values of $Q$ can range from about 0.4 up to 2.0, though the vast majority of measured values in the literature are between 0.6 and 1.0. Low values of the $Q$ parameter ($Q < 0.8$) indicate a clumpy or substructured spatial distribution, while high values ($Q > 0.8$) indicate a smooth, centrally concentrated distribution \citep{Cartwright04}.

\section{Observational data} \label{sec_obs_data}

\subsection{Data collection and data properties} \label{sec_data_colec}

A data set of currently or recently star-forming regions was assembled from three sources. The first is the recent 3D kinematic study of young stellar groups from the Gaia-ESO Survey \citep[][hereafter GES]{Wright24}, with cluster members determined using spectroscopic indicators of youth, such as Lithium absorption or surface gravity indicators. The second source is the APOGEE-2 study of the Orion complex \citep[][hereafter K18]{Kounkel18}, wherein cluster members are identified using a clustering algorithm applied to a sample of previously-identified YSOs. The third is the sample of nearby open clusters from \citet{GaudinAnders20} and \citet{Gaudin20} (hereafter CG20), which applied a clustering algorithm to Gaia DR2 data to identify open clusters and then used a neural network to derive cluster properties. The latter list was limited to clusters with ages $<$ 10 Myrs and distances $<$ 2 kpc to ensure a high kinematic data quality and maximise the chances of detecting residual kinematic structure. 

A final cut was made to the combined list of regions assembled from these three sources; regions with fewer than 30 identified members have been excluded as low-number statistics mean results they produce have reduced reliability. Trumpler 14 is also excluded, as its data appear to suffer from significant contamination from the surrounding Carina association.

A complete list of the 48 regions that were ultimately selected is shown in Table \ref{clust_props} with cluster ages and {\it Gaia} distances taken from each catalogue paper. Absolute stellar ages for young stars are notoriously difficult to estimate accurately, due to uncertainties in both stellar evolutionary models and the stellar properties necessary to compare with such models. Ages for individual star clusters should therefore be used with caution. An additional source, \citet{Franciosini22}, provides an updated age for 25 Orionis. 

\begin{table*} 
	\centering
	\caption{A list of the regions studied in this paper and their properties. Ages and distances are taken from the respective catalogue papers. The $Q$ parameter is calculated in two dimensions. $N$ gives the number of identified members in each cluster, while $N_Q$ gives the number of stars qualifying as members after the removal of outliers (see section \ref{remove_outliers}), which are used in the calculation of Moran's $I$. The estimated contamination and completeness rates are presented as fractions, and the measured $I$($v_{2D}$) of the region is recorded in the final column.}
	\label{clust_props}
\begin{tabular}{lccccccccr}
\hline
Region & Age (Myr) & Distance (Kpc) & $Q$ & $N$ & $N_Q$ & Contamination & Completeness & Source & $I$($v_{2D}$) \\
\hline
22 Orionis & 5.92 & 0.348 $\pm$ 0.004 & 0.73 & 35 & 33 & 0.06 & 0.15 & K18 & 0.03 $\pm$ 0.02 \\
25 Orionis & 19.00 & 0.342 $\pm$ 0.027 & 0.96 & 245 & 199 & 0.06 & 0.15 & GES \& K18 & 0.01 $\pm$ 0.0 \\
ASCC 21 & 8.91 & 0.341 $\pm$ 0.015 & 0.81 & 90 & 87 & 0.05 & 0.25 & CG20 & 0.0 $\pm$ 0.01 \\
Alessi 20 & 9.33 & 0.412 $\pm$ 0.012 & 0.80 & 119 & 118 & 0.05 & 0.25 & CG20 & 0.09 $\pm$ 0.02 \\
BDSB96 & 7.94 & 1.144 $\pm$ 0.131 & 0.82 & 65 & 64 & 0.05 & 0.15 & CG20 & 0.05 $\pm$ 0.02 \\
BH 205 & 6.17 & 1.604 $\pm$ 0.18 & 0.80 & 55 & 55 & 0.05 & 0.15 & CG20 & 0.06 $\pm$ 0.03 \\
Barnard 30 & 2.74 & 0.389 $\pm$ 0.039 & 0.65 & 105 & 88 & 0.06 & 0.15 & GES \& K18 & 0.26 $\pm$ 0.01 \\
Barnard 35 & 2.58 & 0.394 $\pm$ 0.03 & 0.67 & 97 & 81 & 0.06 & 0.15 & GES \& K18 & 0.24 $\pm$ 0.01 \\
Berkeley 87 & 8.32 & 1.644 $\pm$ 0.148 & 0.94 & 112 & 112 & 0.05 & 0.15 & CG20 & 0.01 $\pm$ 0.01 \\
Biurakan 2 & 9.33 & 1.723 $\pm$ 0.156 & 0.82 & 47 & 46 & 0.05 & 0.15 & CG20 & 0.03 $\pm$ 0.01 \\
Bochum 13 & 8.91 & 1.666 $\pm$ 0.143 & 0.88 & 57 & 55 & 0.05 & 0.15 & CG20 & 0.01 $\pm$ 0.02 \\
Christmas Tree Cluster & 2.00 & 0.696 $\pm$ 0.002 & 0.85 & 290 & 195 & 0.05 & 0.15 & GES & 0.03 $\pm$ 0.0 \\
Collinder 197 & 6.00 & 0.981$\pm ^{0.014}_{0.013}$ & 0.81 & 139 & 92 & 0.05 & 0.15 & GES & 0.05 $\pm$ 0.01 \\
Dias 5 & 9.55 & 1.303 $\pm$ 0.13 & 0.76 & 61 & 61 & 0.05 & 0.15 & CG20 & 0.05 $\pm$ 0.02 \\
Epsilon Orionis & 4.33 & 0.365 $\pm$ 0.002 & 0.91 & 70 & 62 & 0.06 & 0.15 & K18 & 0.06 $\pm$ 0.01 \\
Eta Orionis & 5.77 & 0.354 $\pm$ 0.003 & 0.81 & 83 & 83 & 0.06 & 0.15 & K18 & 0.06 $\pm$ 0.0 \\
Gamma Velorum & 19.50 & 0.336 $\pm$ 0.001 & 0.85 & 95 & 95 & 0.05 & 0.15 & GES & -0.01 $\pm$ 0.0 \\
Gulliver 10 & 9.33 & 0.613 $\pm$ 0.029 & 0.71 & 42 & 41 & 0.05 & 0.25 & CG20 & -0.01 $\pm$ 0.03 \\
Gulliver 2 & 6.61 & 1.375 $\pm$ 0.107 & 0.99 & 59 & 57 & 0.05 & 0.15 & CG20 & 0.01 $\pm$ 0.01 \\
IC 1805 & 7.59 & 1.964 $\pm$ 0.178 & 0.97 & 106 & 103 & 0.05 & 0.15 & CG20 & 0.07 $\pm$ 0.01 \\
L1614-N & 2.52 & 0.396 $\pm$ 0.006 & 0.73 & 65 & 65 & 0.06 & 0.15 & K18 & 0.17 $\pm$ 0.01 \\
L1614-S & 1.98 & 0.416 $\pm$ 0.005 & 0.76 & 42 & 41 & 0.06 & 0.15 & K18 & 0.1 $\pm$ 0.02 \\
L1647 & 1.82 & 0.443 $\pm$ 0.007 & 0.63 & 33 & 33 & 0.06 & 0.15 & K18 & 0.18 $\pm$ 0.02 \\
Lambda Orionis & 6.76 & 0.394 $\pm$ 0.028 & 0.97 & 190 & 160 & 0.06 & 0.15 & GES \& K18 & 0.02 $\pm$ 0.0 \\
NGC 1981 & 4.46 & 0.359 $\pm$ 0.003 & 0.74 & 39 & 39 & 0.06 & 0.15 & K18 & 0.09 $\pm$ 0.02 \\
NGC 2244 & 2.00 & 1.401$\pm ^{0.025}_{0.024}$ & 0.72 & 109 & 78 & 0.05 & 0.15 & GES & 0.06 $\pm$ 0.01 \\
NGC 2362 & 5.75 & 1.341 $\pm$ 0.122 & 0.94 & 144 & 139 & 0.05 & 0.15 & CG20 & -0.0 $\pm$ 0.01 \\
NGC 6193 & 5.13 & 1.264 $\pm$ 0.13 & 0.95 & 428 & 423 & 0.05 & 0.15 & CG20 & 0.05 $\pm$ 0.0 \\
NGC 6383 & 3.98 & 1.117 $\pm$ 0.166 & 0.94 & 245 & 242 & 0.05 & 0.15 & CG20 & 0.0 $\pm$ 0.0 \\
NGC 6530 & 1.50 & 1.269$\pm ^{0.035}_{0.033}$ & 0.90 & 639 & 513 & 0.05 & 0.15 & GES & 0.02 $\pm$ 0.0 \\
NGC 6871 & 5.50 & 1.72 $\pm$ 0.137 & 0.84 & 430 & 429 & 0.05 & 0.15 & CG20 & 0.08 $\pm$ 0.0 \\
ONC & 2.51 & 0.398 $\pm$ 0.002 & 0.81 & 670 & 668 & 0.06 & 0.15 & K18 & 0.04 $\pm$ 0.0 \\
Ori C-C & 6.49 & 0.42 $\pm$ 0.003 & 0.85 & 115 & 110 & 0.06 & 0.15 & K18 & 0.05 $\pm$ 0.0 \\
Pismis 27 & 5.37 & 1.883 $\pm$ 0.182 & 0.79 & 35 & 35 & 0.05 & 0.15 & CG20 & 0.06 $\pm$ 0.03 \\
Pismis Moreno 1 & 8.91 & 0.981 $\pm$ 0.067 & 0.86 & 62 & 62 & 0.05 & 0.25 & CG20 & 0.05 $\pm$ 0.01 \\
Pozzo 1 & 9.55 & 0.33 $\pm$ 0.012 & 0.87 & 341 & 341 & 0.05 & 0.25 & CG20 & 0.04 $\pm$ 0.01 \\
Psi 2 Orionis & 5.72 & 0.344 $\pm$ 0.003 & 0.79 & 64 & 64 & 0.06 & 0.15 & K18 & 0.06 $\pm$ 0.01 \\
RCW 33 & 5.00 & 0.926 $\pm$ 0.003 & 0.67 & 192 & 153 & 0.05 & 0.15 & GES & 0.22 $\pm$ 0.01 \\
Sigma Orionis & 2.24 & 0.413 $\pm$ 0.004 & 0.94 & 100 & 91 & 0.06 & 0.15 & K18 & 0.07 $\pm$ 0.01 \\
Spokes Cluster & 1.00 & 0.698 $\pm$ 0.002 & 0.88 & 251 & 190 & 0.05 & 0.15 & GES & 0.09 $\pm$ 0.01 \\
Trumpler 16 & 2.00 & 2.35$\pm ^{0.024}_{0.025}$ & 0.82 & 205 & 149 & 0.05 & 0.15 & GES & 0.01 $\pm$ 0.0 \\
UBC 323 & 8.91 & 1.496 $\pm$ 0.129 & 0.81 & 267 & 267 & 0.05 & 0.15 & CG20 & 0.16 $\pm$ 0.01 \\
UBC 344 & 3.47 & 1.919 $\pm$ 0.181 & 0.83 & 314 & 314 & 0.05 & 0.15 & CG20 & 0.13 $\pm$ 0.01 \\
UPK 402 & 7.24 & 0.433 $\pm$ 0.022 & 0.72 & 47 & 47 & 0.05 & 0.25 & CG20 & 0.02 $\pm$ 0.03 \\
VdBergh 80 & 6.46 & 0.91 $\pm$ 0.089 & 0.85 & 75 & 74 & 0.05 & 0.25 & CG20 & 0.04 $\pm$ 0.01 \\
VdBergh 92 & 7.76 & 1.193 $\pm$ 0.104 & 0.83 & 168 & 166 & 0.05 & 0.15 & CG20 & 0.03 $\pm$ 0.01 \\
Vela OB2 & 14.00 & 0.37 $\pm$ 0.002 & 0.76 & 73 & 59 & 0.05 & 0.15 & GES & 0.1 $\pm$ 0.01 \\
Zeta Orionis & 3.29 & 0.365 $\pm$ 0.005 & 0.84 & 45 & 44 & 0.06 & 0.15 & K18 & 0.06 $\pm$ 0.02 \\
\hline
\end{tabular}

\end{table*}

In addition to name, age, and distance, Table \ref{clust_props} lists a number of other properties. The $Q$ parameter was measured for all clusters, with measured values ranging from approximately 0.6 to 1.0, This is consistent with typical measurements in the literature and indicates a range of spatial morphologies from substructured to centrally-concentrated.

The fraction of the observed stars that are expected to be contaminants is also listed, alongside the estimated completeness of each sample. The contamination of the regions presented in K18 is estimated in that work to be 6 per cent. In the case of the other regions presented here we estimated the contamination of the regions to be 5 per cent, consistent with estimates from astrometric or spectroscopic membership studies \citep{Jackson22}. These estimates are based on the telescope used to gather the data (i.e. $Gaia$), and the distance to the region. Completeness is likewise estimated based on the telescope and the region's properties. Finally, the source of the data for a given region and its calculated Moran's $I$ value, are listed in the final two columns of the table.

Four regions (25 Orionis, Barnard 30, Barnard 35, and Lambda Orionis) occur in both the GES and K18 data sets. In these cases the two data sets are combined to produce a more complete membership list. The age and distance estimates of the two studies are averaged, and stars which appear in both data sets are identified and de-duplicated.  

\subsection{Removing outliers} \label{remove_outliers}

Inspection of the data shows that many clusters contain extreme kinematic outliers. These kinematic outliers may be contaminants or runaway stars, but regardless they are not representative of the kinematic substructure of the region and need to be removed from the sample before it is analysed. We therefore remove all stars with speeds more than 2.5 times the interquartile range (IQR) from the median speed (equivalent to excluding stars more than approximately 3 $\sigma$ from the centre of a Gaussian distribution). The median and IQR are used in the place of the mean and standard deviation as they are less vulnerable to distortion by outliers.

The effects of this cut are evaluated using 20 simulated datasets from A22 that consist of highly substructured, subvirial regions of 1500 stars evolved using the $N$-body integrator \textsc{kira} from the \textsc{starlab} code \citep{Zwart99, Zwart01} to an age of 4 Myr (so some dynamical evolution can take place and the data is more comparable to the observed regions). The $I$($v_{2D}$) of these uncontaminated regions, $I_{\rm{True}}$, is calculated for each simulated region.

Twenty extreme kinematic outliers are then added to each pristine data set, their directions chosen at random and their speeds drawn from a Gaussian distribution with a width ten times that of the pristine data set\footnote{Note that these simulated contaminants are not intended to realistically depict the real contaminants. It is impossible to simulate the distribution of contaminants accurately without identifying them to model their properties, and by definition contaminants are stars which have not been correctly identified. These extreme simulated outliers are used instead as a `sanity check' to check if stars with very obviously deviant kinematics can be removed with reasonable accuracy.}. The $2.5 \times$ IQR cut is then applied to each simulated data set to remove the kinematic outliers. This process was repeated fifty times for each of the twenty simulations. $I$($v_{2D}$) is calculated ($I_{\rm{Calc}}$) for each of the the contaminated and cleaned data sets.

\begin{figure}
	\includegraphics[width=\columnwidth]{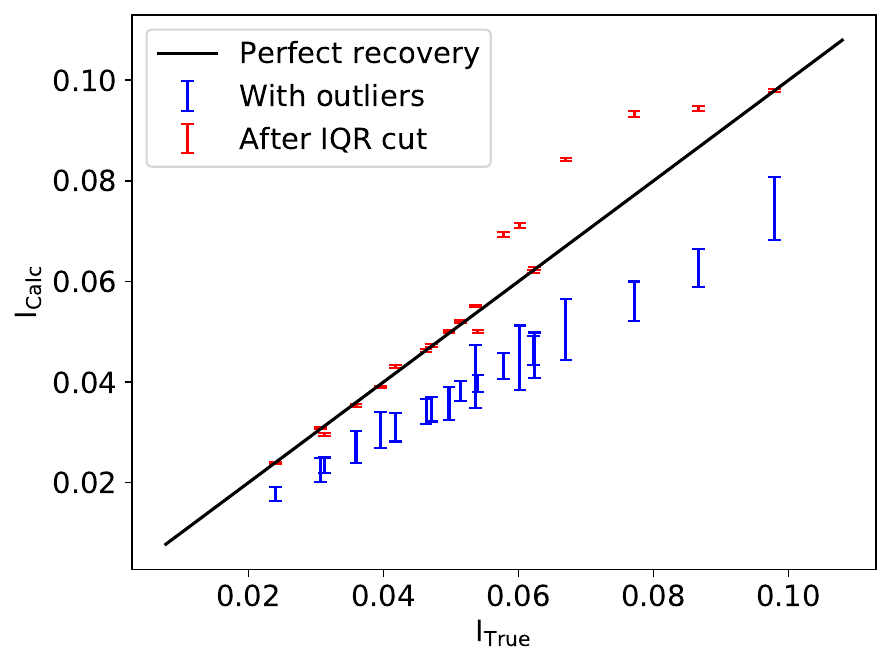}
    \caption{Comparison of the true and calculated Moran's $I$ statistic for data sets containing outliers (blue) and after an IQR cut has been applied with the goal of removing those outliers (red). The true Moran's $I$ statistic values ($I_{\rm{True}}$) are shown on the $x$-axis, and the calculated Moran's $I$ statistic values ($I_{\rm{Calc}}$) are shown on the $y$-axis. A black line is used to indicate where these two values are equal.}
    \label{fig_1}
\end{figure}

Fig. \ref{fig_1} shows how the true values of Moran's $I$ statistic compare to those calculated from the contaminated and cleaned simulations. It is apparent that the effect of adding kinematic outliers is to reduce the measured value of Moran's $I$ statistic (similar to the effects of most observational biases on Moran's $I$, as found by A22). The IQR cut, however, brings most measurements of Moran's $I$ statistic back in agreement with the true value. For some of the regions with a Moran's $I$ statistic $> 0.06$ the IQR cut results in an over-estimation of the value of $I$. We note however that the N-body simulations that result in such over-estimations had all undergone atypical evolution and as a result have unusual spatial and kinematic structures (e.g. one had fragmented into three separate clusters).

The number of genuine and contaminant stars removed by the IQR cut was also recorded from this experiment. On average 12 $\pm$ 11 (0.81 $\pm$ 0.76 per cent) of real stars were removed by the cut, while 17 $\pm$ 2 (86 $\pm$ 9 per cent) of the simulated outliers were removed. These results make it clear that, while the 2.5 IQR cut is not perfect, it does a good job of removing the majority of extreme outliers with relatively little infraction upon the genuine stars. 

More stringent cuts are considered, but are rejected as they would inevitably result in cutting more genuine stars alongside contaminants. This trade off is deemed not worth taking as A22 shows Moran's $I$ can give noisy results at low completeness fractions, but demonstrates high resilience against contamination. More stringent kinematic cuts are also rejected on the grounds that removing kinematic outliers from a study into kinematic structure risks removing the very structure that is being studied. This relatively permissive 2.5 IQR cut was therefore used to remove extreme kinematic outliers from all datasets. These cleaned data sets were then used to produce all the results presented in this paper.

\section{Results} \label{sec_results}

\subsection{$I$($v_{2D}$) of the regions} \label{I_v2D_results}

Following A22, $I$($v_{2D}$) is calculated for each of our observed regions. To estimate the uncertainty on the calculated $I$($v_{2D}$) values, simulated `observed' velocities are drawn for each star. These are drawn from a Gaussian distribution centered on the velocity that was actually observed, with a standard deviation equal to the uncertainty on the velocity measurement for that star. This creates a plausible alternative data set and $I$($v_{2D}$) is recalculated. This process is repeated 1000 times, and the standard deviation of the `observed' $I$($v_{2D}$) values is taken as the uncertainty on the $I$($v_{2D}$) calculated from the original data.

\begin{figure*}
	\includegraphics[width=\textwidth, totalheight=21cm, keepaspectratio]{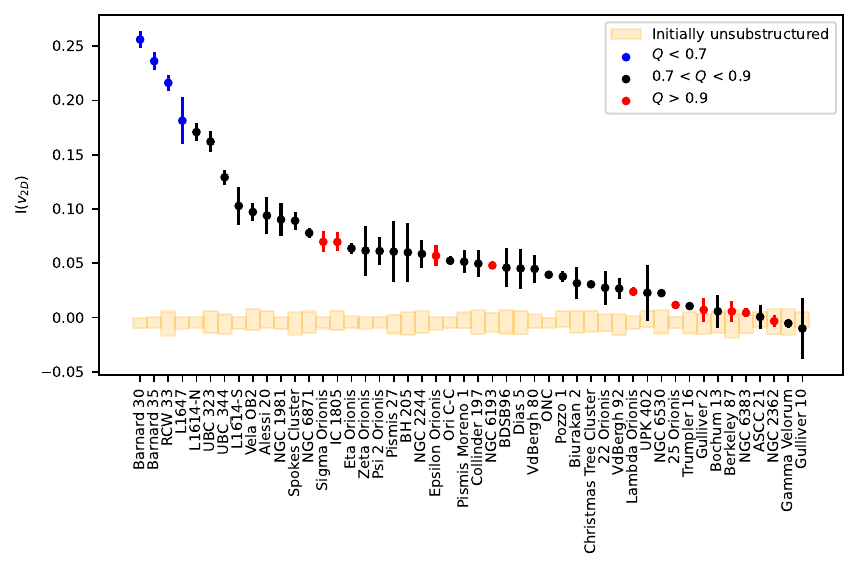}
    \caption{$I$($v_{2D}$) of the observed regions in descending order. The range of expected $I$($v_{2D}$) values for each region, assuming that they formed without initial substructure, is shown in orange. The colour of each point indicates its approximate $Q$ parameter, divided between low ($Q < 0.7$), intermediate ($0.7 < Q < 0.9$) and high ($Q > 0.9$).}
    \label{fig_2}
\end{figure*}

The measured values of $I$($v_{2D}$) and their estimated uncertainties are shown in Fig. \ref{fig_2}, ordered from highest to lowest $I$($v_{2D}$). The range of measured values of $I$($v_{2D}$) from $\sim$0 to $\sim$0.25 is consistent with the predictions from A22 (see the left-hand panels of Figure~2 of that paper) in the hierarchical mergers scenario. These values are larger than those predicted by A22 in the monolithic cluster formation scenario, which typically cover the range of 0.0--0.05, except for older, unbound systems (see the bottom-right panel Figure 2 of A22), which are generally considered to be OB associations and are not well represented in our sample.

\subsection{$I$($v_{2D}$) compared to the $Q$ parameter}

Fig. \ref{fig_2} also shows the approximate $Q$ parameter of each cluster. It is clear that spatially substructured regions $Q < 0.7$ tend to have the highest levels of kinematic substructure, while centrally concentrated spatial distributions ($Q > 0.9$) tend to have low levels of kinematic substructure. A Kolmogorov-Smirnov test \citep{KS_test_ref} confirms that the $I$($v_{2D}$) distributions of regions with $Q < 0.7$ are significantly different from those with $Q > 0.9$ with a $p$-value of 0.002. This shows, as would be expected theoretically, that the $Q$ parameter and $I$($v_{2D}$) are measuring similar aspects of the level of substructure in a region.

\subsection{Comparison with simulation predictions}

Shown in orange on Fig. \ref{fig_2} are the expected range of $I$($v_{2D}$) values for each region if the region formed monolithically. The expected range of values is calculated using the set of twenty simulations without initial kinematic or spatial substructure presented in A22. The calculation is performed using snapshots taken at the age of the relevant region\footnote{For the two regions that are older than 10~Myr, the maximum age of the $N$-body simulations in A22, we use the simulation snapshots at 10 Myr, as the vast majority of the dynamical evolution of the region has already occurred by this time and very little further change is expected beyond 10~Myr.}. Measurement uncertainty, contamination and incompleteness are all incorporated, as described in A22, at the levels listed in Table~\ref{clust_props}.

From inspection of Fig. \ref{fig_2} it is apparent that for 39 out of the 48 regions studied, $I$($v_{2D}$) is at least one standard deviation above the range expected if the regions formed in a manner consistent with the monolithic model, but is consistent with formation by hierarchical mergers. In the case of five of the remaining nine regions (UPK 402, Gulliver 2, Bochum 13, Berkeley 87, and NGC 6383) $I$($v_{2D}$) is above the shaded orange area, but the error bar extends down into the monolithic-compatible zone. This indicates they most likely formed with kinematic structure, but could plausibly have formed without substructure. The error bars for ASCC 21 and Gulliver 10 extend above, throughout and below that zone, meaning they are potentially compatible with having formed with kinematic structure, without kinematic structure, and with kinematic anti-structure. It is therefore determined that no conclusions can be drawn from the structure of those regions. The data points and error bars for the remaining two regions (Gamma Velorum, and NGC 2362) exist solely within the orange region, indicating they show no signs of kinematic substructure.

Gamma Velorum has an estimated age of 19.5 Myr \citep{Jeffries17}, and given its relatively compact nature at this age \citep[its radius is 1.9 pc,][]{Wright24}, it is presumed to be bound. These two factors mean that the region has most likely undergone a great deal of dynamical evolution, so it is to be expected that no signature of initial kinematic structure could be recovered (even if it was initially present). No conclusions can be meaningfully drawn from the non-detection of significant structure in Gamma Velorum.   

NGC 2362 does not have any obvious features which would explain its low $I$($v_{2D}$). However, A22 demonstrated that simulated regions initialised with statistically identical levels of kinematic structure evolve stochastically, resulting in a wide range in $I$($v_{2D}$) values. After being evolved to 5.75 Myr (the age of NGC 2362) the observed values of $I$($v_{2D}$) in these simulations are, by a small margin, potentially compatible with the assumption of no initial substructure. This means it is plausible that NGC 2362 did form with kinematic substructure despite that structure's present absence. Of course, the possibility also remains that NGC 2362 did not form with kinematic structure. 

The remaining 39 of the 48 regions have $I$($v_{2D}$) values significantly in excess of what would be expected had the regions been initially unsubstructured. Notably there are six regions whose $Q$ parameter implies a circular and centrally-concentrated morphology ($Q > 0.9$), while their $I$($v_{2D}$) values imply that they have significant levels of kinematic substructure. These regions are Sigma Orionis, IC~1805, Epsilon Orionis, NGC 6193, Lambda Orionis and 25 Orionis. The majority of these are well-known and quite compact young (Sigma Ori) or open clusters (NGC 6193 and Lambda Ori), the latter of which are old and compact enough to very likely be gravitationally bound. Only Epsilon Ori is young and sparse enough to potentially be gravitationally unbound, while IC~1805 is suspected to have an expanding halo that dominates its kinematic substructure (see Section \ref{sec_expand}). \footnote{We note that this makes it difficult to determine if the high $I$($v_{2D}$) found in IC~1805 is attributable to primordial structure of developed expansion structure or, most probably, both.}

In A22 it was shown that spatially clustered but kinematically substructured conditions are incompatible with having formed via the monolithic model of star cluster formation, but are compatible with the hierarchical model. This observational evidence provides a direct constraint on star formation models; star forming regions must, and least predominantly, produce stars with spatial-kinematic substructure. Further, these must be formed in such a manner that this structure can survive the erasing effects of dynamical evolution for at least 5 and up to 10+ Myr.   

\subsection{Alternative methods of quantifying kinematics} \label{alt_quant}

In section \ref{method_morans} several additional kinematic parameterisations of Moran's $I$ beyond $I$($v_{2D}$) are outlined. These metrics are calculated for each of the observed regions, and the regions are ranked from highest to lowest Moran's $I$ according to each metric. From those results it is clear that the Moran's $I$ values of different kinematic parameters are generally similarly ranked for a given region. 

This reinforces the key finding of section \ref{I_v2D_results}; kinematic structure is present in the vast majority of the observed regions. If this were not the case, and the identified structure was due to chance alignments, then it is highly unlikely structure would be present to similar extent in transverse velocity components, and the other presented paramaterisations. See appendix \ref{append_A} for more details of this analysis and Figure \ref{fig_append_A} for a comparison of the rankings of the different metrics.

\subsection{A closer look at individual regions}

The kinematic structure of regions of particular interest is now discussed. To aid the interpretation of this structure, in this analysis the LISA statistic is applied to each region. Recall from section \ref{method_LISA} that the LISA statistic quantifies the degree to which each data point contributes to the substructure observed in the region. In this section the regions are plotted with the stars and their velocity vectors colour coded according to their LISA statistic, to highlight the most significant kinematic substructures. 

\subsubsection{Expanding regions} \label{sec_expand}

\begin{figure*}
	\includegraphics[width=\textwidth, totalheight=21cm, keepaspectratio]{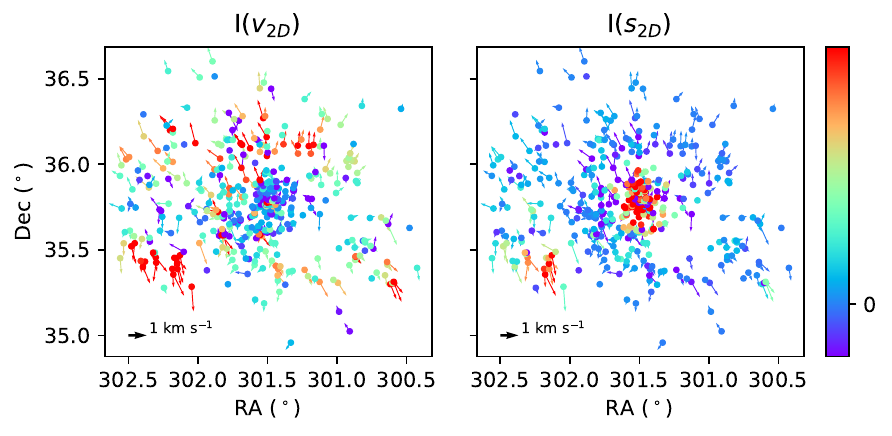}
    \caption{A plot of NGC 6871 colour coded by the stellar LISA values of $I$($v_{2D}$) (left) and $I$($s_{2D}$) (right). To make the velocity vectors of the stars more easily interpretable, a vector showing 1 km s$^{-1}$ is shown in black.}
    \label{fig_3}
\end{figure*}

Several regions with high $I$($v_{2D}$) have clear visual indications of expansion in their spatial morphology and kinematics, notably NGC 6871, Barnard 35, RCW 33, UBC 323, UBC 344 and IC 1805. NGC 6871 is depicted in Fig. \ref{fig_3}. In the left hand panel of this figure the stars are colour coded according to the LISA statistic of $I$($v_{2D}$), and in the right hand panel the colour coding is done  according to the the LISA statistic of $I$($s_{2D}$). Comparable plots for the other five regions can be found in appendix \ref{expand_append}. In these figures the velocity vectors are in km s$^{-1}$ but re-scaled uniformly such that the vectors fit well within the boundaries of the plot, rather than being too large/small to be easily interpreted. This is done the make the kinematic structure of the regions, which is the focus of this paper, as apparent as possible.

From the left hand panels of these six figures, it is evident that the stars at the largest radii tend to make the most significant contributions to the kinematic structure as quantified by $I$($v_{2D}$). These stars are predominately those with the strongest radial trajectories, which is why they are found at such large distances from the centre (assuming they have expanded ballistically). This indicates that expansion is the central feature driving the kinematic substructure of these regions.

In the right hand panels of these six figures the stars are colour coded by their LISA values of $I$($s_{2D}$) instead of $I$($v_{2D}$). In each of these regions it is strikingly apparent that highly significant structure is present in their cores. This is particularly true for the four oldest regions, RCW 33, NGC 6871, UBC 323, and IC 1805, all of which have ages $\geq$ 5 Myr. The cores appear divorced from their host regions kinematically, in a very harsh, precipitous manner. This is in contrast to the comparatively gentle transition in their spatial structure from high density to low density. Note that while the right hand panels of these figures highlight that these core regions contain stars with similar speeds, these stars are not identifiable in the left hand panels as having structure in their velocities. There is also a fraction (which varies in size between these regions) of stars which are not highlighted in either figure, indicating they have no kinematic substructure whatsoever. These primarily reside at intermediate radii between the core and the expanding edges.

The most plausible explanation for this three-tiered structure (core, unsubstructured halo, expanding fringe) is that stars with sufficient initial kinetic energy to escape their host region's potential well did so very quickly. They then make up the stars driving the observed expansion, as has already been discussed. In contrast, stars with insufficient kinetic energy to escape formed a dense, bound core. Due to their low kinetic energy these core stars have similar speeds (close to zero in the rest frame in the region), so are identified as a spatial-kinematic substructure by LISA of $s_{2D}$. Here the rest frame is taken as the median stellar velocity of the observed members. The frequent dynamical interactions inevitable in such a bound core continually randomise the directions of the star's velocities, accounting for the lack of structure in $v_{2D}$ apparent there. The intermediate region, i.e. the unsubstructured halo, is most likely made up of stars whose initial velocities fell on the borderline of the regions' escape velocity at early times. These stars have likely continued to mix in the halo, accounting for the lack of structure in $v_{2D}$ and $s_{2D}$. 

\subsubsection{NGC 6530}

\begin{figure*}
\includegraphics[width=\textwidth]{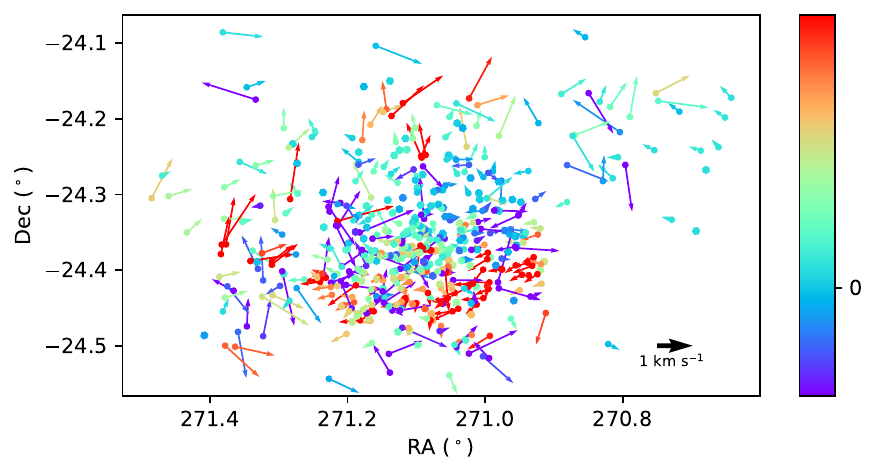}
    \caption{NGC 6530 with stellar velocity vectors colour coded according to their LISA values of $I$($v_{2D}$). To make the velocity vectors of the stars more easily interpretable, a vector showing 1 km s$^{-1}$ is shown in black.}
    \label{fig_4}
\end{figure*}

NGC 6530 is an extremely young region with an estimated age of just 1.5 Myr. It is also large and relatively well sampled, with 513 observed members that pass our quality cuts. NGC 6530 has highly significant substructure as quantified by $I$($v_{2D}$) and shown in Fig. \ref{fig_4}. Examination of this figure shows what appears to be an outward moving spiral structure of stars. This structure begins at an RA of 270.95$^{\circ}$ and Dec of -24.40$^{\circ}$, before proceeding anticlockwise by approximately 180$^{\circ}$, ending at around RA $=$ 271.00$^{\circ}$, Dec $=$ -24.15$^{\circ}$. Other observable properties of this outwards spiral structure are that stellar density appears to decrease along its length, and stellar speeds appear to increase.

The `spiral' is investigated further to verify whether it is a single, cohesive stream as it appears. This is done by producing a colour magnitude diagram of apparent members of the `spiral', as well as examining their parallaxes and kinematics. From this it is determined that this structure is in fact a chance alignment of four sub-regions, rather than a cohesive stellar stream. The existence of these identified kinematic substructures within NGC 6530 re-enforces the main finding of section \ref{I_v2D_results}; young regions contain significant spatial-kinematic substructure. This is in keeping with the hierarchical model of star formation, and in opposition to the monolithic model.

%\begin{figure}
%\includegraphics[width=\columnwidth]{fig_8.pdf}
%    \caption{The four groups identified in the apparent spiral structure in NGC 6530.}
%    \label{fig_8}
%\end{figure}

\section{Discussion and conclusions} \label{sec_conclusions}

Spatial-kinematic data for 48 young clusters and star forming regions has been analysed. Moran's $I$ statistic is used to quantify the degree of kinematic structure in each region, and the results are assessed in the context of the predictions made by A22, which are:

\begin{itemize}
    \item If star clusters form according to the monolithic model their kinematic structure, as quantified by $I$($v_{2D}$), is expected to be zero within a very small, region-dependent margin of error.
    \item If star clusters form according to the hierarchical model their kinematic structure, as quantified by $I$($v_{2D}$), is expected to be significantly above zero for ages up to and potentially beyond 8 Myr. At older ages $I$($v_{2D}$) is expected to occupy a significantly enlarged range of possible values compared to the monolithic model. 
\end{itemize}

The regions observed are predominantly (39 out of 48 regions) found to have highly significant levels of kinematic substructure, to a degree that is incompatible with those regions having formed according to the monolithic model of star cluster formation. The results for the remaining nine regions are potentially ambiguous. The relationship between spatial and kinematic substructure is investigated by comparing Moran's $I$ and the $Q$ parameter. It is found that there is a strong correlation between the presence and strength of spatial and kinematic substructure, but that there are multiple clusters that lack spatial substructure, but have kinematic substructure. This is also predicted by, and therefore supports, the hierarchical model of star cluster formation. We conclude that on the whole the observational evidence demonstrates that stars are, at least predominantly, born with a high degree of kinematic substructure, as expected by the hierarchical mergers model of star cluster formation. Even if the hierarchical model is not assumed, star formation models must be able to account for the early and enduring kinematic substructure observed in these regions.

\section*{Acknowledgements}

NJW and BA acknowledge a Leverhulme Trust Research Project Grant (RPG-2019-379), which supported this research.

%%%%%%%%%%%%%%%%%%%%%%%%%%%%%%%%%%%%%%%%%%%%%%%%%%
\section*{Data Availability}

The data analysed in this paper is publicly available on VizieR. The sample from GES is presented in \citet{Wright24}. The data from K18 can be found at \url{https://doi.org/10.26093/cds/vizier.51560084} and the data for CG20 can be found at \url{https://doi.org/10.26093/cds/vizier.36400001}.

%%%%%%%%%%%%%%%%%%%% REFERENCES %%%%%%%%%%%%%%%%%%

% The best way to enter references is to use BibTeX:

\bibliographystyle{mnras}
\bibliography{complete_manuscript_file} 

\appendix 

\section{Alternative methods of quantifying kinematics, detailed results} \label{append_A}

As described in section \ref{alt_quant} Moran's $I$ of a variety of different kinematic parameterisation is calculated and ranked for each of the 48 regions discussed in this paper. In some regions certain paramaterisations cannot be calculated due to a sparsity or complete absence of radial velocity data. In that case those regions are not included in the ranking for the relevant metrics.

\begin{figure*}
\centering
	\includegraphics[width=\textwidth, totalheight=21cm, keepaspectratio]{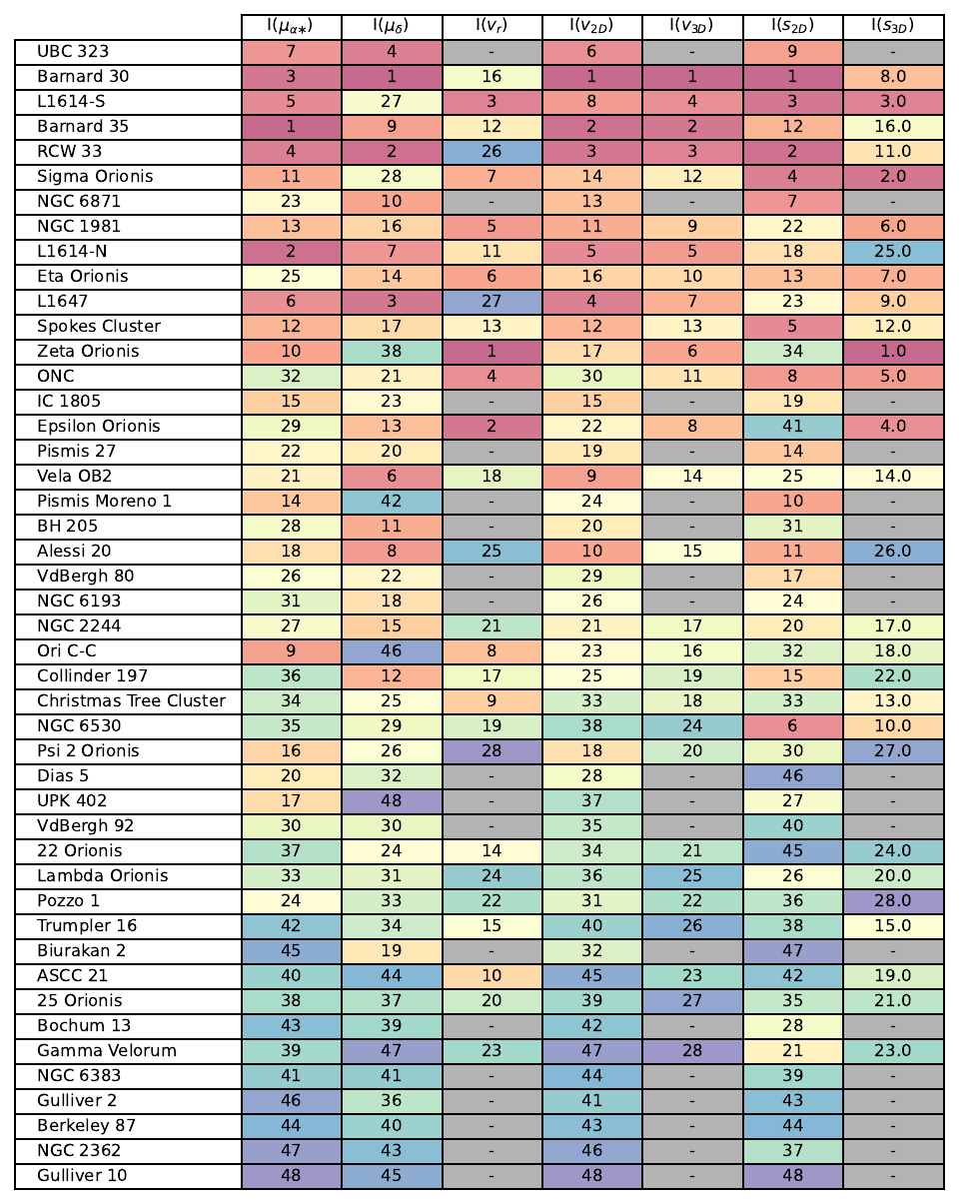}
    \caption{The ranks of each region's Moran's $I$ values according to different kinematic metrics, colour coded by rank.}
    \label{fig_append_A}
\end{figure*}

The results of the ranking are presented in Fig. \ref{fig_append_A}. This figure contains a cell diagram in which each cell is colour coded according to the rank it contains, in order to make trends more obvious. The regions are listed in this figure in order of highest to lowest relative mean rank, i.e. the average rank accounting for the absence of some rankings due to missing radial velocity data. There is a relatively smooth gradient from red to blue which can be observed in the figure. This indicates that a given region tends to have similar kinematic structure rankings across the different parameterisations, as outlined in section \ref{alt_quant}.

The metric with the least smooth descending gradient in Moran's $I$ is $v_r$. This is likely due to the impact of binary stars, whose radial velocities can distort the observed distribution. It may also be a result of the realtive sparsity of radial velocity data in some regions. In such regions $I$($v_{r}$) is calculated from fewer data points than $I$($\mu_{{\alpha}{\ast}}$), $I$($\mu_{\delta}$), and their derivatives, so it suffers from a higher level of noise. 

\section{Additional expanding regions}\label{expand_append}

\begin{figure*}
\centering
	\includegraphics[width=\textwidth, totalheight=21cm, keepaspectratio]{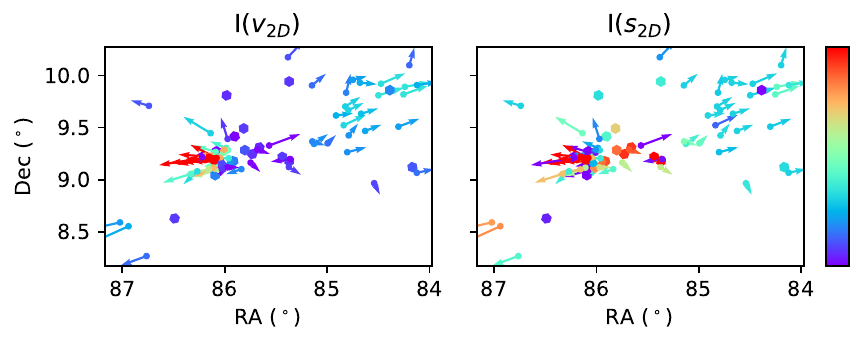}
    \caption{A plot of Barnard 35 colour coded by the stellar LISA values of $I$($v_{2D}$) (left) and $I$($s_{2D}$) (right).}
    \label{fig_append_b1}
\end{figure*}

\begin{figure*}
\centering
	\includegraphics[width=\textwidth, totalheight=21cm, keepaspectratio]{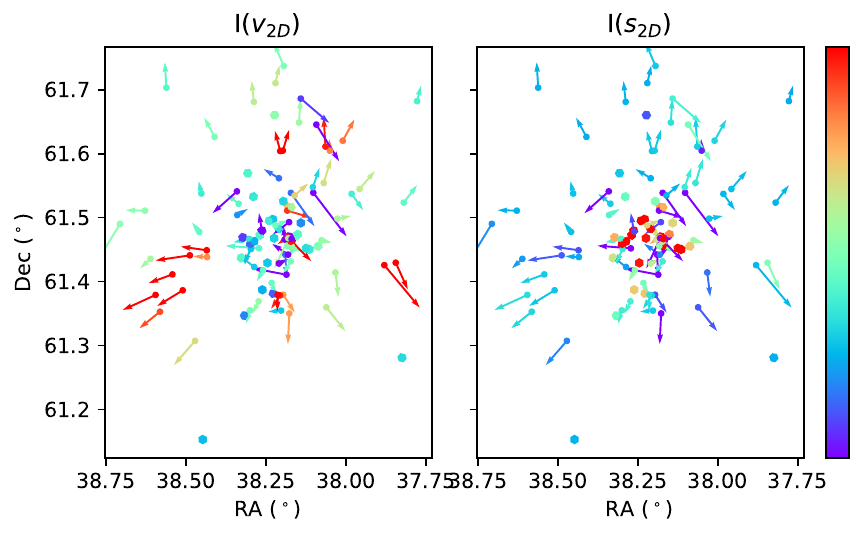}
    \caption{A plot of IC 1805 colour coded by the stellar LISA values of $I$($v_{2D}$) (left) and $I$($s_{2D}$) (right).}
    \label{fig_append_b2}
\end{figure*}

\begin{figure*}
\centering
	\includegraphics[width=\textwidth, totalheight=21cm, keepaspectratio]{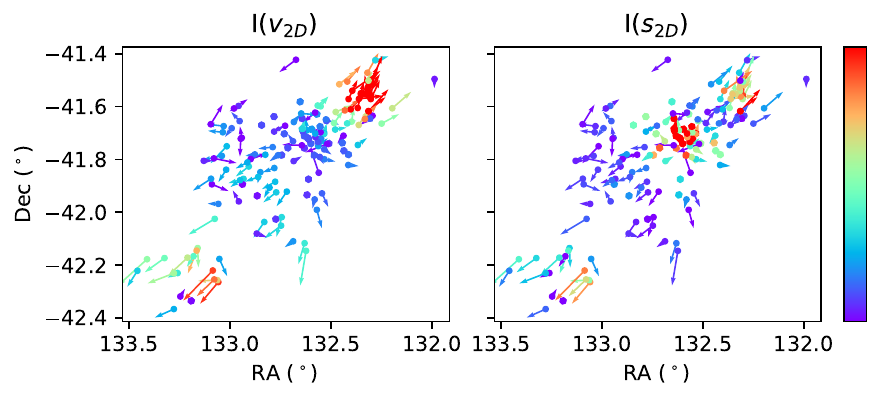}
    \caption{A plot of RCW 33 colour coded by the stellar LISA values of $I$($v_{2D}$) (left) and $I$($s_{2D}$) (right).}
    \label{fig_append_b3}
\end{figure*}

\begin{figure*}
\centering
	\includegraphics[width=\textwidth, totalheight=21cm, keepaspectratio]{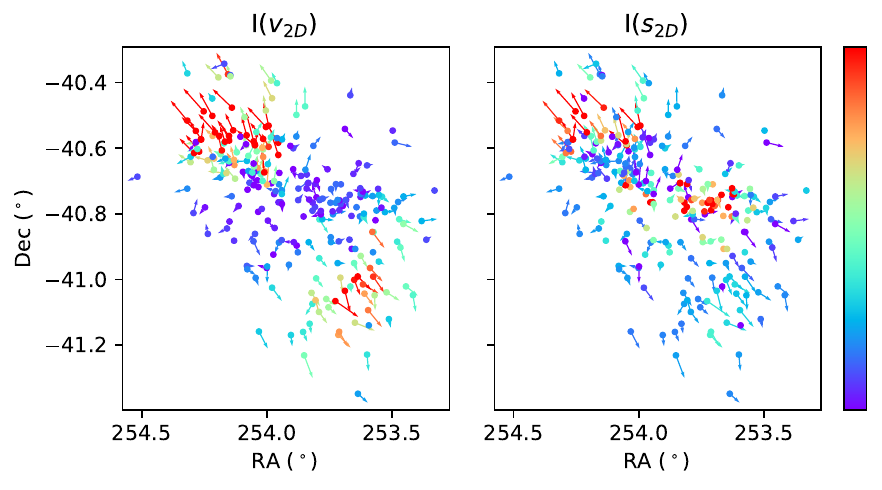}
    \caption{A plot of UBC 323 colour coded by the stellar LISA values of $I$($v_{2D}$) (left) and $I$($s_{2D}$) (right).}
    \label{fig_append_b4}
\end{figure*}

\begin{figure*}
\centering
	\includegraphics[width=\textwidth, totalheight=21cm, keepaspectratio]{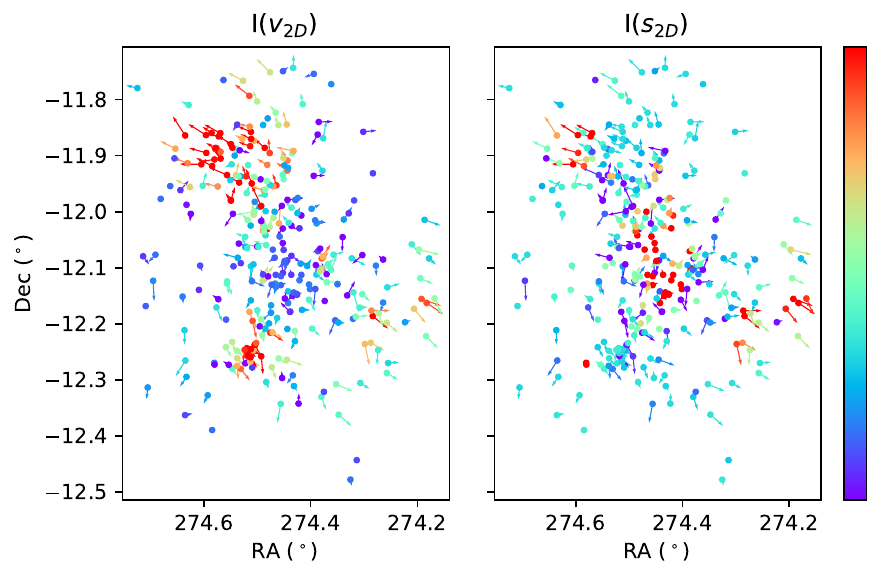}
    \caption{A plot of UBC 344 colour coded by the stellar LISA values of $I$($v_{2D}$) (left) and $I$($s_{2D}$) (right).}
    \label{fig_append_b5}
\end{figure*}

This appendix shows the LISA plots of additional expanding regions. Barnard 35 is shown in Fig. \ref{fig_append_b1}, IC 1805 is shown in Fig. \ref{fig_append_b2}, RCW 33 in Fig. \ref{fig_append_b3}, UBC 323 in Fig. \ref{fig_append_b4}, and UBC 344 in Fig. \ref{fig_append_b5}.

%%%%%%%%%%%%%%%%%%%%%%%%%%%%%%%%%%%%%%%%%%%%%%%%%%

%%%%%%%%%%%%%%%%%%%%%%%%%%%%%%%%%%%%%%%%%%%%%%%%%%

% Don't change these lines
\bsp	% typesetting comment
\label{lastpage}
\end{document}